%% file: main.tex
\documentclass[10pt, conference, a4paper]{IEEEtran}
\usepackage{algorithmic}
\usepackage{textcomp}
\usepackage{multirow}
\usepackage{eurosym}
\usepackage{color, soul}
\usepackage{graphicx,epsfig,epstopdf,amsmath,amssymb, amsfonts,enumerate}
\usepackage{url}
\usepackage{geometry}

\begin{document}

\newgeometry{left=1.39cm, right=1.39cm, top=1.9cm, bottom=4.29cm}
\title{Cost Analysis for Drone based 5G eMBB Provision to Emergency Services} 

\author{
\IEEEauthorblockN{Mythri Hunukumbure}
\IEEEauthorblockA{\textit{Samsung Research United Kingdom}\\
Staines-Upon-Thames - UK\\
mythri.h@samsung.com}
\and
\IEEEauthorblockN{Galini Tsoukaneri}
\IEEEauthorblockA{\textit{Samsung Research United Kingdom}\\
Staines-Upon-Thames - UK\\
g.tsoukaneri@samsung.com}
}

\maketitle

\begin{abstract}
There is growing interest in investigating how the 5G capabilities can be extended to the critical communications vertical, in an expedited time scale. This paper provides a comprehensive cost analysis of a drone-based 5G deployment solution for emergency services. The proposed deployment uses existing 4G and 5G ground small cells and the fronthaul, backhaul network. The cost analysis develops an analytical approach to minimize the \textit{Total Cost of Ownership (TCO)}, where the key parameters of drone unit cost, number of drones per link, and link capacity are jointly optimized. Our numerical analyses look at the TCO sensitivity to the above parameters to find \textit{sweet spots} for deployment in different parameter combinations. The 3GPP centralization options are also investigated in terms of the TCO, and it is observed that centralization split points at higher layers benefit from lower TCO in this deployment model.     
\end{abstract}

\begin{IEEEkeywords}
5G, Centralized RAN, Emergency services, Drones, CAPEX, OPEX, TCO, Techno-economic analysis, 
\end{IEEEkeywords}

\input{Introduction}
\input{DeploymentModel}

\input{AnalyticalModel}

\input{Evaluation}

\input{Conclusions}

\small
\bibliographystyle{IEEEtran}
\bibliography{citations}

\end{document}

%% file: Introduction.tex
\section{Introduction}
\label{sec:intro}

There is a concerted effort in many countries to switch the emergency communications from specialized networks (e.g. TETRA) to wider cellular networks. The ability of the cellular networks to handle larger traffic volumes in major events, seamlessly support mobile IP-based services, while allowing for cost and efficiency savings are the main reasons for this switch. The \textit{Emergency Services Network (ESN)} in the UK~\cite{esn}, and FirstNet in the US~\cite{frn} are prime examples of this transformation, with both deployments expected t run on 4G LTE infrastructure, with additional quality and reliability guarantees. Within this context, it is logical to look at what 5G can offer, and how to provide early 5G access to these emergency communication networks.

5G promises to provide increased data rates, similar to both \textit{enhanced Mobile Broadband (eMBB)} and \textit{Ultra-Reliability Low Latency Communications (URLLC)}, which can be highly useful in the emergency communications domain~\cite{Markakis:2017}. Downloading interactive 3D maps on the move to an emergency response location and uploading \textit{Ultra-High Definition (UHD)} videos from inside the locations are some of the potential eMBB applications. The URLLC capabilities can be utilized to operate robots remotely in potentially hazardous conditions, for example.

A key question is how to overcome the localized nature of early 5G networks to extend the connectivity throughout an emergency service area (i.e. an area affected by a natural disasters, fire, etc.) In this paper, we propose a drone-based 5G eMBB service, where wireless links are employed to connect the drones to a suitably provisioned 5G and 4G ground small cell network, and subsequently to the emergency command/control center through the network core. In this deployment, we look at the main cost drivers, and develop an analytical optimization model to minimize the TCO of deploying and operating such a service. Additionally, we conduct a detailed numerical cost analysis, considering two centralized RAN options proposed by 3GPP~\cite{3gpp.38.801}. Finally, we provide a numerical analysis on the sensitivity of the TCO to the main cost drivers identified in the study (i.e. number of drones, drone unit costs, capacity increases), giving indications about the best deployment model under certain constraints.

The rest of the paper is organized as follows. In Section~\ref{sec:deploModel}, we present the deployment model and identify its key cost components. In Section~\ref{sec:mathAnalysis} we present the analytical TCO optimization against key cost components identifies in Section~\ref{sec:deploModel}. In Section~\ref{sec:evaluation} we present our numerical results for different TCO variations on the centralized RAN options, and TCO sensitivity to key parameters. We conclude in section~\ref{sec:conclusions} and discuss planned further work.

%% file: DeploymentModel.tex
\section{Deployment Model}
\label{sec:deploModel}

The deployment model is based on using the drones in a flexible manner, to provide 5G eMBB connectivity to the emergency crew on the scene. This will be an \textit{on demand} service, where the emergency service can request this 5G link based on the particular needs and the severity of the emergency. The key requirement is that the 5G connectivity can be deployed to any location within the coverage area of the particular emergency service with the quality of the connectivity guaranteed. In this study, we assume the coverage area to be within a metropolitan city limits.

\begin{figure}
  \centering
    \includegraphics[width=0.45\textwidth]{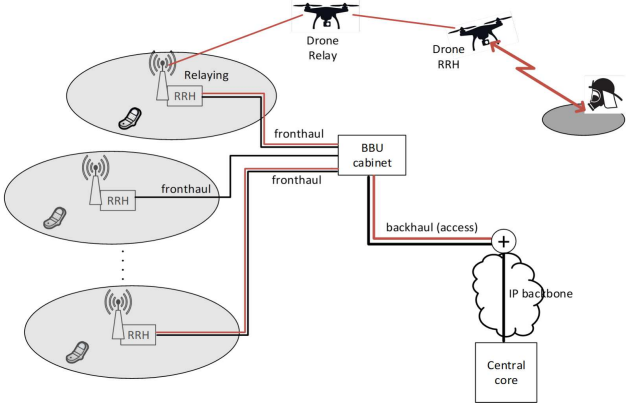}
  \caption{\textbf{Drone-based deployment for emergency service support.}}
  \label{fig:droneService}
\end{figure}

The deployment model considered in this study is illustrated in Fig.~\ref{fig:droneService}. We assume the use of \textit{Radio Access Network (RAN)} centralization (CRAN), according to which, drones are equipped with \textit{Remote Radio Heads (RRHs)} with the minimum radio kit required for data transmission, and \textit{Base Band Unit (BBU)} processing is carried out at a central unit. To support the drone link, a number of pre-selected ground small cells needs to be upgraded to relay the signal to/from the drones to the BBU processor, and the core network. Thus, \textit{additional} capacity needs to be provisioned both in the fronthaul and backhaul networks to support this service.

The number of drones used in the wireless link and the small cells needed to support the drone connectivity offer a tradeoff point in the analysis. With a higher number of drones, the reach of the wireless link increases and, thus, fewer ground small cells need to be upgraded and provisioned for additional fronthaul capacity. We assume a good spread of 4G and 5G small cells in the city, so any configuration of drone numbers in the wireless link can be supported by upgrading existing ground small cells. Furthermore, the BBU are assumed to have enough capacity to include new connections from the drone cells.

In this study, the main cost points include the drone unit costs, the costs for upgrading the small cells, and the incremental costs for providing additional fronthaul and backhaul capacity. These individual cost components will vary according to the number of drones considered for the wireless link, the RAN centralization options and the capacity provisioned
in the drone link, as will be seen in the TCO evaluations in the following sections.

%% file: AnalyticalModel.tex
\section{Analytical Model for TCO Minimization}
\label{sec:mathAnalysis}
In this section we present a mathematical analysis to minimize the TCO of the drone-based deployment for emergency service support.

\subsection{Mathematical Analysis}
In this analysis, our objective is to minimize the TCO of the drone service, which is a function of the number of drones $n_{dr}$, the capacity increment $c$, and the unit cost $d$ of each drone (Sec.~\ref{sec:deploModel}). The capacity increment essentially determines the additional capacity supported by the RRHs of the drones, as well as the additional capacity that needs to be supported in the fronthaul and backhaul. As such, we define the capacity increment as: $c = c_{base} * (c_{step}-1) * 100$, where $c_{base}$ is the base channel capacity with $c_{base} = 665$ Mbps, and $ci$ is the capacity step with $c_{step} = \left\{1, 2, ..., m \right\}$.

Based on our drone deployment model (Sec.~\ref{sec:deploModel}), we define the total TCO as:

\begin{equation}
	TCO = C_{dr} + C_{sc} + C_{fh} + C_{bh}
\label{eq:tco}
\end{equation}

where $C_{dr}$ are the drone-related costs (e.g. drone cost, drone RRHs costs), $C_{sc}$ is the total cost to upgrade the selected ground small cells, $C_{fh}$ is the fronthaul cost for the increased capacity, and $C_{bh}$ is the backhaul cost for the additional capacity required.

The $C_{dr}$ is defined as:

\begin{equation}
	C_{dr} = 2^{c_{step}-1} * n_{dr} * d
\label{eq:droneCost}
\end{equation}

The $C_{sc}$ is defined as:
\begin{equation}
	C_{sc} = n_{sc} * n_{dr} * smc
\label{eq:scCost}
\end{equation}
where $smc$ is the upgrade cost of each small cell, and $n_{sc}$ is the number of ground small cells that need to be upgraded to support the drone link, and is equal to: 
\begin{equation}
	n_{sc} = \frac{city\_area}{\pi * (0.2*n_{dr})^2} = k*\frac{1}{n_{dr}^2}
\label{eq:smNumber}
\end{equation}


The $C_{fh}$ is defined as:
\begin{equation}
	C_{fh} = n_{sc}*[a*\left(fhc +n_{dr}*c\right)^{b} - a*fhc^{b}]
\label{eq:fhCost}
\end{equation}
where $fhc$ is the fronthaul capacity, and $a$, $b$ are network operator specific cost parameters.

Finally, the $C_{bh}$ is defined as:
\begin{equation}
	C_{bh} = bbu*\left[a*\left(\frac{bhc + n_{dr}*c*k}{mux}\right)^b - a*\left(\frac{bhc}{mux}\right)^b\right]
\label{eq:bhCost}
\end{equation}
where $bbu$ is the number of BBUs in the city area, $bhc$ is the backhaul capacity, and $mux$ is the multiplexing gain, with $mux = 1.5$.

Based on the above definitions, our objective function to minimize the TCO is defined as:
\begin{align}
f(n_{dr}, c_{step}) &= 2^{c_{step}-1} * n_{dr} * d +  \frac{k}{n_{dr}} * smc + \nonumber \\
         &\qquad {}  \frac{k}{n_{dr}^2} * a * [\left(fhc +n_{dr}*c\right)^{b} - fhc^{b}]+ \nonumber \\
         &\qquad {}  bbu*a*\left[\left(\frac{bhc + c*\frac{k}{n_{dr}}}{mux}\right)^b - \left(\frac{bhc}{mux}\right)^b\right]
         \label{eq:objFun}
\end{align}
based on the constraints $c_1: n_{dr} \geq 1$ and $c_2: c_{step} \geq 1$. Please note that the unit cost $d$ of each drone only affects the $C_{dr}$ in a linear and monotonic function. As such, the minimum $C_{dr}$ is achieved with with minimum value for $d$, and therefore, we do not further include it in our optimization analysis.

As this is a multi-coordinate optimization problem (considering the $n_{dr}$ and $c_{step}$ as our coordinates), we use the coordinate gradient descent approach~\cite{Tseng:2009} to minimize along the multiple coordinate directions. To do so, we first need to find the partial derivative functions over the $n_{dr}$ and $c_{step}$ coordinates. Based on~\ref{eq:objFun}, the partial derivative function $f'$ over $n_{dr}$ is:
\begin{align}
f'(n_{dr}) &= 2^{c_{step}-1}  * d + \frac{1}{n_{dr}^2} * k*smc + \nonumber \\
         &\qquad {} k * a \left\{ \frac{2}{n_{dr}^3} * \left( fhc + n_{dr}*c\right)^b + \right.\nonumber\\ 
         &\qquad \left. \frac{1}{n_{dr}^2} * \left[ b* \left( fhc + n_{dr}*c\right)^{b-1} \right] +  \right.\nonumber\\ 
         &\qquad \left. fhc^b * \frac{2}{n_{dr}^3}\right\} +  \nonumber \\
         &\qquad {} bbu * a* b * \left( \frac{bhc*n_{dr} + c*k}{n_{dr}*mux}\right)^{b-1} 
\label{eq:derivN}
\end{align}

Similarly, the partial derivative function $f'$ over $c_{step}$ is:
\begin{align}
f'(c_{step}) &= n_{dr} * d * (c_{step}-1) * 2^{c_{step}-2} + \nonumber \\
         &\qquad {} k*a*\frac{1}{n_{dr}^2 } * \left[ b* \left(fhc + n_{dr}*c \right)^{b-1} \right] + \nonumber \\
         &\qquad {} bbu * a * \frac{1}{\left( n_{dr}*mux\right)^b} * \nonumber \\
         &\qquad {} \left[ b * \left( bhc*n_{dr} + c* k\right)^{b-1} \right] 
\label{eq:derivC}
\end{align}

Therefore, the gradient descent with respect to each coordindate can be determined based on:

\begin{equation}
\nabla f = \begin{bmatrix}
			eq. 8\\
			eq. 9\\
		   \end{bmatrix}
\label{eq:gradientDesc}
\end{equation}
subject to the constraints:
\begin{equation}
\begin{bmatrix}
		\frac{\vartheta c_1}{\vartheta n_{dr}}		& \frac{\vartheta c_2}{\vartheta n_{dr}}\\
		\frac{\vartheta c_1}{\vartheta c_{step}}	& \frac{\vartheta c_2}{\vartheta c_{step}}\\
\end{bmatrix} = 
\begin{bmatrix}
		1	& -c_{step}\\
		1	& -n_{dr}\\
\end{bmatrix}
\label{eq:constraints}
\end{equation}

\begin{figure}
  \centering
    \includegraphics[width=0.45\textwidth]{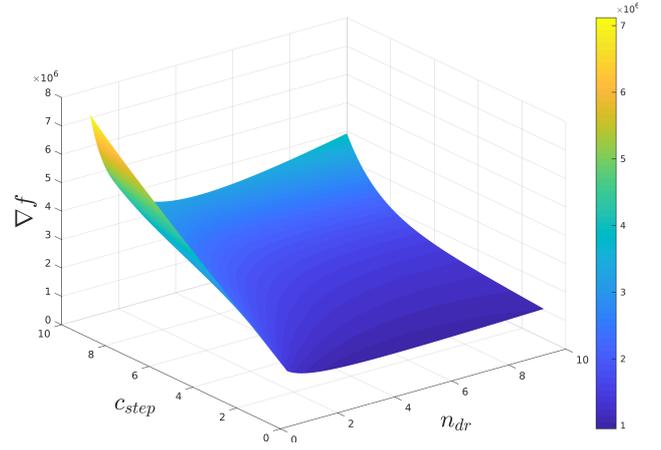}
  \caption{\textbf{Gradient descent for $\nabla f$.}}
  \label{fig:gradDesc}
\end{figure}

For our gradient descent analysis we assumed the constants as defined in Table~\ref{tab:constants}.
Figure~\ref{fig:gradDesc} shows the gradient descent of Eq.~\ref{eq:gradientDesc} for $n_{dr} = \left\{ 1, ..., 10 \right\}$, $c_{step} = \left\{ 1, ..., 10 \right\}$, and $step = 0.1$, subject to the constrains of Eq.~\ref{eq:constraints}. Our results show that the TCO is minimized when $n_{dr} = 7$ and $c_{step} = 1$.

\begin{table}
	\centering
	\begin{tabular}{l | p {1.5cm} }
	\hline
	\hline
	\textbf{Constant}				& \textbf{Value}		\\
	\hline
	\hline
	City Area						& $100 km^2$			\\
	\hline
	Drone cost ($d$)				& 9120 \euro			\\
	\hline
	Cost parameter $a$				& $3840$				\\
	\hline
	Cost parameter $b$				& $0.2$					\\
	\hline
	$smc$							& $2550$ \euro			\\
	\hline
	$fch$ 							& $799$	Mbps			\\
	\hline
	$bch$ (per cell)				& $833$	Mbps			\\
	\hline
	$bbu$							& $6$					\\
	\hline
	$mux$							& $1.5$					\\
	\hline
	\end{tabular}
	\vspace{1mm}
	\caption{Gradient descent evaluation constants.}
	\label{tab:constants}
	\vspace{-2mm}
\end{table}

%% file: Evaluation.tex
\section{Evaluation}
\label{sec:evaluation}

With the analytical TCO optimization in place, we now look at a number of numerical evaluations under different conditions. For the numerical values, we utilize fairly accurate
estimates of the costs related to ground small cells, BBU and the fronthaul/backhaul leasing from the mmMAGIC EU project~\cite{mmMagic}. The costs related to the drone units are mostly assumed, and we evaluate the TCO for a wide range of these values, to cover most of the possibilities. It should be emphasized that the trends seen in these numerical analysis are more reliable than the absolute TCO values, which are mostly indicative to the specific deployment conditions we have set forth.

\subsection{TCO variations for centralized options}
\begin{figure}
  \centering
    \includegraphics[width=0.45\textwidth]{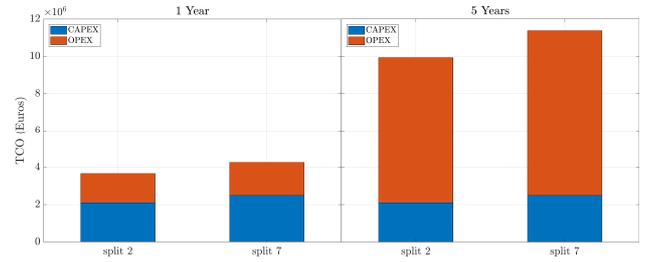}
  \caption{\textbf{CAPEX and OPEX costs of the proposed scheme for the first and fifth year of operation and the two splits considered.}}
  \label{fig:tcoAll}
\end{figure}

We consider two centralization options as specified by the 3GPP~\cite{3gpp.38.801}, the CRAN split option $7$ which is at the PHY layer, and the split option $2$ which is at the PDCP layer. With split $7$, the drone radio unit needs to function only as an RF transmitter, so its costs will be lower. However for this split, the fronthaul data rates need to be higher which, in turn, increments the fronthaul costs, as well as the costs related to ground small cell upgrades. For split $2$, the processing up to the PDCP layer happens in the drone radio unit, thus the opposite cost factors are effective in this split. The $1$ year and $5$ year TCO variations for the two split options are shown in Figure~\ref{fig:tcoAll}.

Our results show that the CRAN split $2$ (PDCP layer) produces lower costs both in the $1$ year and $5$ year TCO. This split has lower OPEX in terms of reduced fronthaul capacity and leasing costs. The results also indicate that the OPEX costs play a major role in determining the TCO for this high capacity, 5G eMBB based connectivity solution.

\subsection{TCO sensitivity to drone unit cost and the number of drones per wireless link}

One of the key parameters studied in the analytical model in section~\ref{sec:mathAnalysis} is the optimum number of drones per wireless link ($n_{dr}$). This number also varies with the unit cost of the drone, which was fixed to a cost of the small cell RRH in section~\ref{sec:mathAnalysis}. Here, we evaluate the TCO with both the drone unit cost ($d$), and the number of drones per wireless link ($n_{dr}$) as variables. We consider the TCO related to the CRAN split $2$, as this split produces the lower TCO.

\begin{figure}
  \centering
    \includegraphics[width=0.45\textwidth]{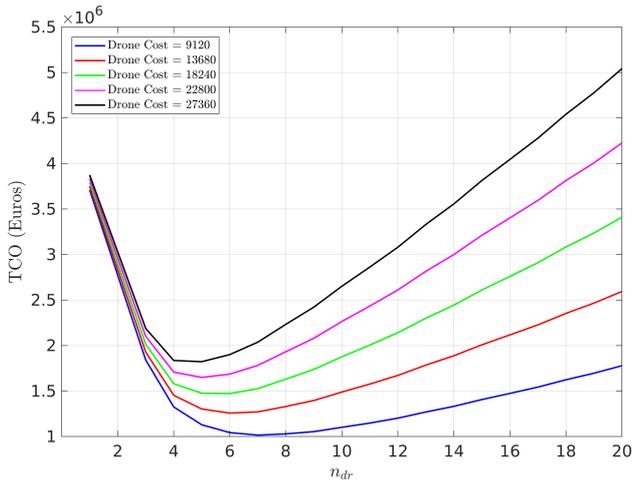}
  \caption{\textbf{Experimental TCO for variable  number of drones per wireless link ($n_{dr}$) and different drone unit costs, assuming a fixed capacity increment of $c_{step}=1$.}}
  \label{fig:costAnalysis}
\end{figure}

The drone unit cost is varied in steps of $4000$\euro \hspace{0.9mm} from the minimum cost considered in~\ref{sec:mathAnalysis}. Please note that the TCO includes the same main cost components as discussed in the previous section. The results are shown in Figure~\ref{fig:costAnalysis}.

The results indicate that as the drone unit cost increases, the minimum TCO point moves to lower numbers of $n_{dr}$. This is quite logical, as higher drone costs will offset the
savings on the reduced number of small cell upgrades given by higher $n_{dr}$. The minimum TCO variation also indicates that this \textit{sweet spot} for the TCO will have to be carefully analyzed before the actual deployments take place.

\subsection{TCO sensitivity to capacity increment}

Another key cost parameter in this study is the required capacity that must be provisioned in the wireless drone link. To provide true eMBB services to multiple users, we estimate that a minimum of 665 Mbps link capacity ($c_{base}$) should be provided per drone link. This is for the consumption in both the uplink and downlink, with the use of dynamic TDD~\cite{eIMTA}. The analytical approach in section~\ref{sec:mathAnalysis} returned the minimum TCO for this minimum capacity. Here, we extend this minimum capacity in $100$ Mbps steps to assess the TCO variations, as certain 5G applications would require even higher capacity from the wireless link.

\begin{figure}[t]
  \centering
    \includegraphics[width=0.45\textwidth]{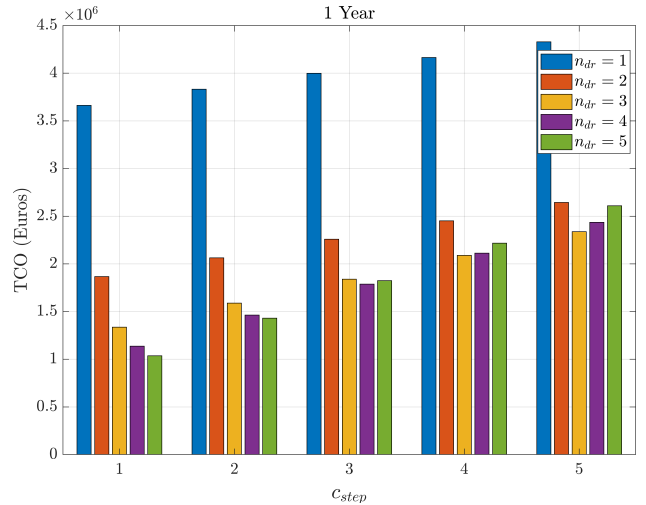}
  \caption{\textbf{Estimated TCO for different $c_{step}$ and $n_{dr}$ values in year $1$.}}
  \label{fig:tcoYear1CI}
\end{figure}

\begin{figure}[t]
  \centering
    \includegraphics[width=0.45\textwidth]{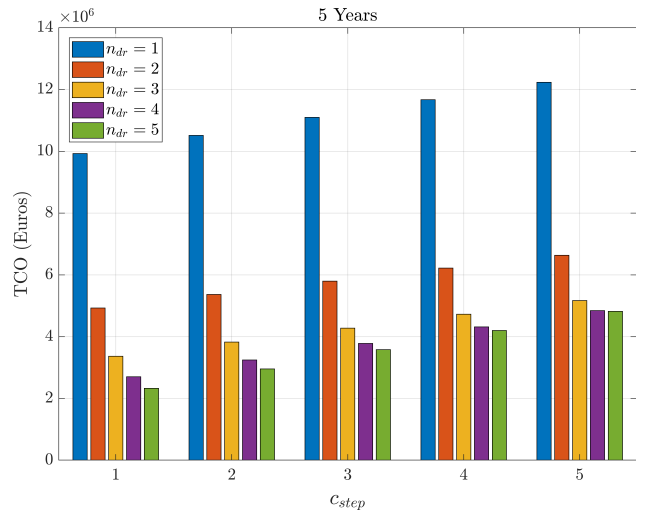}
  \caption{\textbf{Estimated TCO for different $c_{step}$ and $n_{dr}$ values in year $5$.}}
  \label{fig:tcoYear5CI}
\end{figure}

In this study we keep the spectrum allocation for this \textit{on demand} service fixed to 100 MHz. (Details of a separate paper on the spectrum analysis is provided in section~\ref{sec:conclusions}.) Specifically, the capacity is incremented in $100$ Mbps steps by increasing the received SNR in $3 dB$ steps based on the Shannon capacity limits~\cite{commSystems}, as studied in the analytical model (Sec.~\ref{sec:mathAnalysis}). Again, we consider the lower TCO split $2$ option for the analysis. The results for the $1$ year and $5$ year TCO variations for different capacity levels and different $n_{dr}$ is shown in Figures~\ref{fig:tcoYear1CI} and~\ref{fig:tcoYear5CI} respectively.

Our results show that there are two main factors that determine the TCO, when incrementing the capacity. The CAPEX cost for the drone unit doubles with each capacity increment step, in line with more antenna elements, or better RF power amplifiers. The OPEX costs for provisioning fronthaul and backhaul also increase. The relative significance of these costs can be inferred by comparing the $1$ year and $5$ year TCOs in Figures~\ref{fig:tcoYear1CI} and~\ref{fig:tcoYear5CI}. In Figure~\ref{fig:tcoYear1CI}, the CAPEX component of the drone unit cost is significant, and for higher capacity provisions the \textit{sweet spot} of the lowest TCO attains progressively with lower number of drones per link. These shorter drone links will incur higher OPEX costs in fronthaul provision for a higher number of small cells. However as the TCO contains only the OPEX for $1$ year, this increase is dwarfed by the CAPEX savings coming from the lower number of drones per wireless link. With the $5$ year TCO, the fronthaul OPEX costs (accumulated for $5$ years) are dominant, and when increasing the capacities, the lowest TCO point remains with higher number of drones. However, the TCO gains when moving to higher numbers of drones are diminishing with higher capacities, which indicates the counter balancing behavior of the increasing CAPEX costs of drone units.

%% file: Conclusions.tex
\section{Conclusions}
\label{sec:conclusions}

The work presented in this paper looks at the deployment costs of a drone-based 5G eMBB provision for emergency services. Different configurations in terms of the centralization options, the number drones per wireless link, the drone unit costs, and the capacity levels have been analyzed in order to find the trends in TCO variations. Due to the high capacity requirements and thus the higher OPEX costs of fronthaul provision, the higher layer split point in centralization produces lower TCO. With higher drone unit costs (be it for higher precision/reliability or higher capacity provision) the lower TCO point occurs with lower number of drones per link, particularly for the $1$ year TCO. For $5$ year TCO, the fronthaul OPEX costs become dominant, and when providing higher capacities it makes more sense to deploy more drones per wireless links in order to reduce the number of small cells requiring fronthaul capacity increments. The key message in all these analyses is that comprehensive cost estimates for different configurations should be done at the system design level, in order to determine the best option.

The mode of spectrum allocation for these \textit{on demand} wireless links is a key consideration, which we address in a parallel paper~\cite{spatioTempPaper}. In terms of further work, it is planned to present this study to emergency services and receive their feedback to improve some of the assumptions made. Also, this study will be taken to the 5G VINNI EU project~\cite{5gVinni}, to investigate the possibility of developing a basic proof of concept trial using one of their 5G platforms.

\section*{Acknowledgements}
This work has been performed in the framework of the Horizon 2020 project ONE5G (ICT-760809) receiving funds from the European Union. The authors would like to acknowledge the contributions of their colleagues in the project, although the views expressed in this contribution are those of the authors and do not necessarily represent the project.